\documentclass{article}
\usepackage{spconf,amsmath,graphicx,hyperref}
\usepackage{cite}
\usepackage{amssymb,amsfonts}
\usepackage{graphicx}
\usepackage{textcomp}
\usepackage{xcolor}
\usepackage{algorithm}
\usepackage{algpseudocode}
\usepackage{booktabs}
\usepackage[table]{xcolor}
\usepackage{tabularx}
\usepackage{array}
\usepackage{soul} 
\usepackage[most]{tcolorbox}
\usepackage{booktabs,multirow}

\newcolumntype{Y}{>{\centering\arraybackslash}X}

\definecolor{stageblue}{RGB}{220,240,245}
\definecolor{stageorange}{RGB}{245,230,205}
\definecolor{stagepurple}{RGB}{225,230,254}

\newtcolorbox{stagebox}[1]{
    colback=#1,
    colframe=#1,
    boxrule=0pt,
    arc=0pt,
    left=2pt,
    right=2pt,
    top=1pt,
    bottom=1pt,
    boxsep=0pt,
    before skip=1pt,
    after skip=1pt,
    fontupper=\raggedright
}

\newcommand{\method}{\textbf{Aura}} 

\newcommand{\N}{\mathcal{N}}
\newcommand{\bx}{\mathbf{x}}
\newcommand{\by}{\mathbf{y}}
\newcommand{\bz}{\mathbf{z}}

\newcommand{\bn}{\mathbf{n}}
\newcommand{\bX}{\mathbf{X}}
\newcommand{\bR}{\mathbf{R}}

\newcommand{\bW}{\mathbf{W}}

\newcommand{\oA}{\mathcal{A}}
\newcommand{\oD}{\mathcal{D}}
\newcommand{\oM}{\mathcal{M}}

\newcommand{\bI}{\mathbf{I}}

\newcommand{\bv}{\mathbf{v}}

\newcommand{\beps}{\mathbf{\epsilon}}

\DeclareMathOperator*{\argmin}{argmin}
\DeclareMathOperator*{\argmax}{argmax}

\def\BibTeX{{\rm B\kern-.05em{\sc i\kern-.025em b}\kern-.08em
    T\kern-.1667em\lower.7ex\hbox{E}\kern-.125emX}}

\title{Controllable Mono-to-Stereo Upmixing using Reward-tilted Flows}
\name{
Sattwik Basu$^{1^\dagger}$\thanks{\textsuperscript{$\dagger$}This work was done during an internship at Dolby Laboratories, Inc.},
Gautam Bhattacharya$^{2}$,
Alan Seefeldt$^{2}$
}
\address{
$^{1}$University of Illinois Urbana-Champaign, 
$^{2}$Dolby Laboratories, Inc.
}

\begin{document}
\ninept
\maketitle

\begin{abstract}
We consider the problem of mono-to-stereo audio upmixing: generating a high-fidelity two-channel stereo signal from a single-channel recording. 
The task is fundamentally ill-posed i.e., many stereo signals are consistent with a given mono recording.
An effective upmixer must therefore generate a realistic, musically coherent stereo image while preserving the underlying audio content. 
Previous approaches have relied on signal-processing driven decorrelation, ambience extraction and  supervised learning, offering limited spatial control or requiring task-specific training. 
In response, we propose {\method} \textit{(\underline{A}udio \underline{U}pmixing using \underline{R}ew\underline{a}rd-tilted Flows)}, which reformulates upmixing as a generative inverse problem using a pretrained flow prior. 
To enable controllable spatial imaging, we design a reward-tilted posterior sampler whose reward penalizes divergences between generated and target distributions of inter-channel cues. 
Composing forward operators further enables joint restoration and upmixing without retraining or fine-tuning.
Objective and subjective evaluations on real-world music recordings from MUSDB18-HQ demonstrate our method's effectiveness relative to several baselines. 

\end{abstract}

\section{Introduction}
\label{sec:Intro}
Mono-to-stereo upmixing arises naturally in many scenarios: enhancing archival music, restoring legacy film soundtracks, and remixing contemporary recordings captured with a single microphone. 
Converting these recordings to stereo can enrich our listening experience by enhancing spatial separation and ambience, while providing a crucial foundation for surround audio generation.
Unfortunately, mono-to-stereo generation is fundamentally underdetermined: mono audio lacks necessary inter-channel cues needed to uniquely determine source panning or ambient spatial structure.
A convincing upmix must therefore preserve the recorded content while \emph{inventing} perceptually realistic and musically coherent spatial information. 
This challenge is further compounded when the mono recording is degraded by bandwidth limitations, compression artifacts, or nonlinear distortion.

While stereo-to-surround upmixing has received considerable attention, mono-to-stereo remains relatively less explored. 
Existing approaches broadly fall into signal-processing, end-to-end generative, and parametric methods. 
Classical signal-processing techniques \cite{kendall1995decorrelation,fink2015downmix, schlecht2018optimized, alary2017velvet} synthesize inter-channel differences through time delays, decorrelation, multiband processing, or source--ambient decomposition \cite{he2015ambient,breebart}. 
Data-driven approaches learn richer mono-to-stereo mappings. 
DiffStereo \cite{Zhang2025DiffStereoEM}, for example, trains a conditional diffusion model \cite{ho2020denoising} to generate stereo directly from mono observations, requiring task-specific training on mono--stereo pairs. 
Accommodating observation processes beyond those seen during training, such as additional degradations, requires adapting \cite{hu2021lora} or retraining the model. 
Another line of attack aims to predict intermediate stereo parameters \cite{serra}, providing interpretable control while restricting the output to the chosen parameterization.
In summary, these methods are constrained by hand-designed spatial rules, task-specific training, or restricted output parameterizations.

\textbf{This paper} breaks away from prior approaches and reformulates mono-to-stereo upmixing as a generative inverse problem. 
Briefly, we seek to recover an unknown stereo signal $\bx$ from a mono recording $\by=\mathcal{A}(\bx)+\bn$, where $\mathcal{A}$ models a downmixing process and $\bn$ denotes measurement noise. 
As mono recordings lack inter-channel spatial information, $\oA$ has a large null space: many distinct stereo signals can produce the same mono observation. 
Consequently, recovering $\bx$ is severely ill-posed~\cite{hadamard} and requires a strong prior over realistic stereo audio.
Our method, {\method}, uses a pretrained flow model as this prior $p(\bx)$ and casts reconstruction as Bayesian posterior
sampling from $p(\bx|\by)\propto p(\by|\bx)p(\bx)$ \cite{dps}.
Here, the likelihood $p(\by|\bx)$ steers generation towards consistency with the mono observation, while the prior favors realistic high-fidelity stereo audio.
Composing the downmixing operator with a degradation operator further enables joint restoration and upmixing using the same
prior, without task-specific training or fine-tuning.
Moreover, the posterior accommodates multiple plausible stereo reconstructions. 
This provides scope for shaping the stereo image at inference-time. 
We exploit this flexibility through spatial rewards based on divergences between inter-channel cue statistics. 
This effectively tilts the posterior toward desired spatial characteristics while retaining realistic audio content and measurement consistency.

\textbf{Our contributions} are threefold:
(1) a \emph{training-free} formulation for mono-to-stereo upmixing and joint restoration using a pretrained flow prior;
(2) designing \emph{spatial rewards} based on inter-channel coherence and level-difference distributions for controllable stereo generation without retraining;
and (3) a sampling procedure with \emph{decoupled refinement} that separates measurement-consistency from spatial reward optimization.
We evaluate {\method} on music recordings from MUSDB18-HQ~\cite{musdb18-hq}, considering both clean mono downmixes and observations degraded by bandwidth loss, clipping, and additive noise. 
Our evaluation combines objective measures of reconstruction accuracy, spectral fidelity, and spatial realism
with subjective listening tests of perceptual quality and stereo imaging. 
Results demonstrate improvements over strong signal-processing, parametric, and learned generative baselines.
We believe our method generalizes beyond mono-to-stereo upmixing and discuss early thoughts on surround audio generation.
(\textbf{Audio demos} provided \href{https://basusattwik.github.io/aura-demo/}{here.})

\section{Formulation and Background}
\label{sec:Formulation_background}

\textbf{Our formulation} assumes a measurement model $\by = \oA(\bx) + \bn$, where $\bx \in \mathbb{R}^{n \times 2}$ is the unknown stereo signal,
$\by \in \mathbb{R}^{n \times 1}$ is the observed, potentially degraded mono signal, and $\bn \sim \N(\boldsymbol{0},\sigma^2\bI)$ denotes measurement noise.
The forward operator $\oA=\oD\circ\oM:\mathbb{R}^{n\times 2}\to\mathbb{R}^{n\times 1}$ combines stereo-to-mono downmixing $\oM$ with an optional degradation operator $\oD$.
This formulation extends to latent-space models \cite{ldm} with $\bx=\varphi(\bz)$ for decoder $\varphi$; for simplicity, we retain signal-space notation.
The operators $\oD$ and $\oM$ discard crucial spatial and harmonic information, making recovery ill-posed and necessitating a prior over plausible stereo audio.
Classical priors (sparsity, low-rankness, or parametric models \cite{EnglHankeNeubauer1996}) are often restrictive in capturing the complexity of natural audio. 
We instead use a pretrained flow model to provide an expressive data-driven prior $p(\bx)$ over realistic stereo signals.

\textbf{Flow models} use a learned time-dependent velocity field $\bv_\theta(\bx,t)$ to transport samples from a base distribution $p_1$, typically $\N(\boldsymbol{0},\bI)$, to the data distribution $p_0$ through the ODE
$d\bx_t=\bv_\theta(\bx_t,t)\,dt$ \cite{lipman2022flow,liu2022flow}.
The induced density evolution satisfies the continuity equation
$\partial_t p_t(\bx)+\nabla_{\bx}\!\cdot\!\left(p_t(\bx)\bv_\theta(\bx_t,t)\right)=0$ \cite{Batchelor_2000}.
To make training tractable, an interpolating probability path is defined as $\bx_t=(1-t)\bx_0+t\beps$, where $\bx_0\sim p_0$ and
$\beps\sim\N(\boldsymbol{0},\bI)$, yielding the conditional velocity
$\bv_t(\bx_t|\bx_0)=\beps-\bx_0$ \cite{liu2022flow, albergo2025stochastic}.
The model is then trained with the conditional flow-matching objective \cite{lipman2022flow}
\begin{align}
\min_{\theta}\;
\mathbb{E}_{t,\bx_0,\bx_t}
\left[
\left\|
\bv_\theta(\bx_t,t)-\bv_t(\bx_t|\bx_0)
\right\|_2^2
\right].
\label{eq:fm-loss}
\end{align}
At inference, samples are initialized from $\bx_1\sim p_1$ and numerically integrate the learned ODE
backward from $t=1$ to $t=0$ until reaching $\bx_0 \sim p_0$.
This learned flow $\bv_\theta(\bx_t,t)$ implicitly defines a prior $p(\bx)$ over the data as evident from its connection to the score-function \cite{score, kim2025flowdps, ddpm} as given below
\begin{align}
t\,\nabla_{\bx_t}\log p_t(\bx_t) \approx -\bx_t-(1-t)\bv_\theta(\bx_t,t),
\end{align}
This prior regularizes ill-posed inverse problems at inference time, without requiring further task-specific retraining.\cite{dps,kim2025flowdps,inversebench}.

\section{Method}
\label{sec:Method}

We design {\method} as a posterior sampling framework centered on solving two tasks: 1) combining the flow prior with the measurement likelihood
to reconstruct stereo audio, 2) introducing spatial rewards in a computationally tractable manner for enabling controllable stereo imaging at inference-time.

\subsection{Posterior Sampling with Flow Priors}
\label{subsec:posterior_sampling}
Our first task seeks to draw plausible stereo samples from $p(\bx_0|\by)$.
Starting from a noisy state $\bx_t$, we progressively anneal the noise level through a sequence of clean-signal inference and independent
renoising steps \cite{daps, song2019generative}. Each such update is equivalent to drawing $\bx_s\sim p(\bx_s\mid\bx_t,\by)$ at a lower noise level $s<t$, where
\begin{align}
p(\bx_s|\bx_t,\by)
&=
\int p(\bx_s|\bx_t,\by,\bx_0)\,
p(\bx_0|\bx_t,\by)\,d\bx_0
\nonumber\\
&=
\int q_s(\bx_s|\bx_0)\,
p(\bx_0|\bx_t,\by)\,d\bx_0.
\label{eq:daps-transition}
\end{align}
The second equality follows from independent renoising: conditioned on $\bx_0$, the new state $\bx_s$ is independent of $\bx_t$ and $\by$. 
Here, $q_s(\bx_s|\bx_0)$ is the noising kernel associated with the flow model's linear interpolation.
Using Bayes' rule, the posterior in Eq. \ref{eq:daps-transition} is
\begin{align}
p(\bx_0|\bx_t,\by)
\propto
p(\by|\bx_0)
p(\bx_0|\bx_t)
\label{eq:daps-clean-posterior}
\end{align}
where $p(\by|\bx_0) = \mathcal{N}(\by; \oA(\bx_0), \sigma^2 \bI)$ is the likelihood needed to enforce measurement consistency.
We approximate the prior-induced clean-signal distribution as $p(\bx_0|\bx_t)\approx\mathcal{N}(\bx_0;\hat{\bx}_0(\bx_t),r_t^2\bI)$,
centered at the flow's Tweedie estimate~\cite{efron2011tweedie} $\hat{\bx}_0(\bx_t)=\bx_t-t\bv_\theta(\bx_t,t)$.
\begin{figure}[t]
    \centering
    \includegraphics[width=0.99\columnwidth]{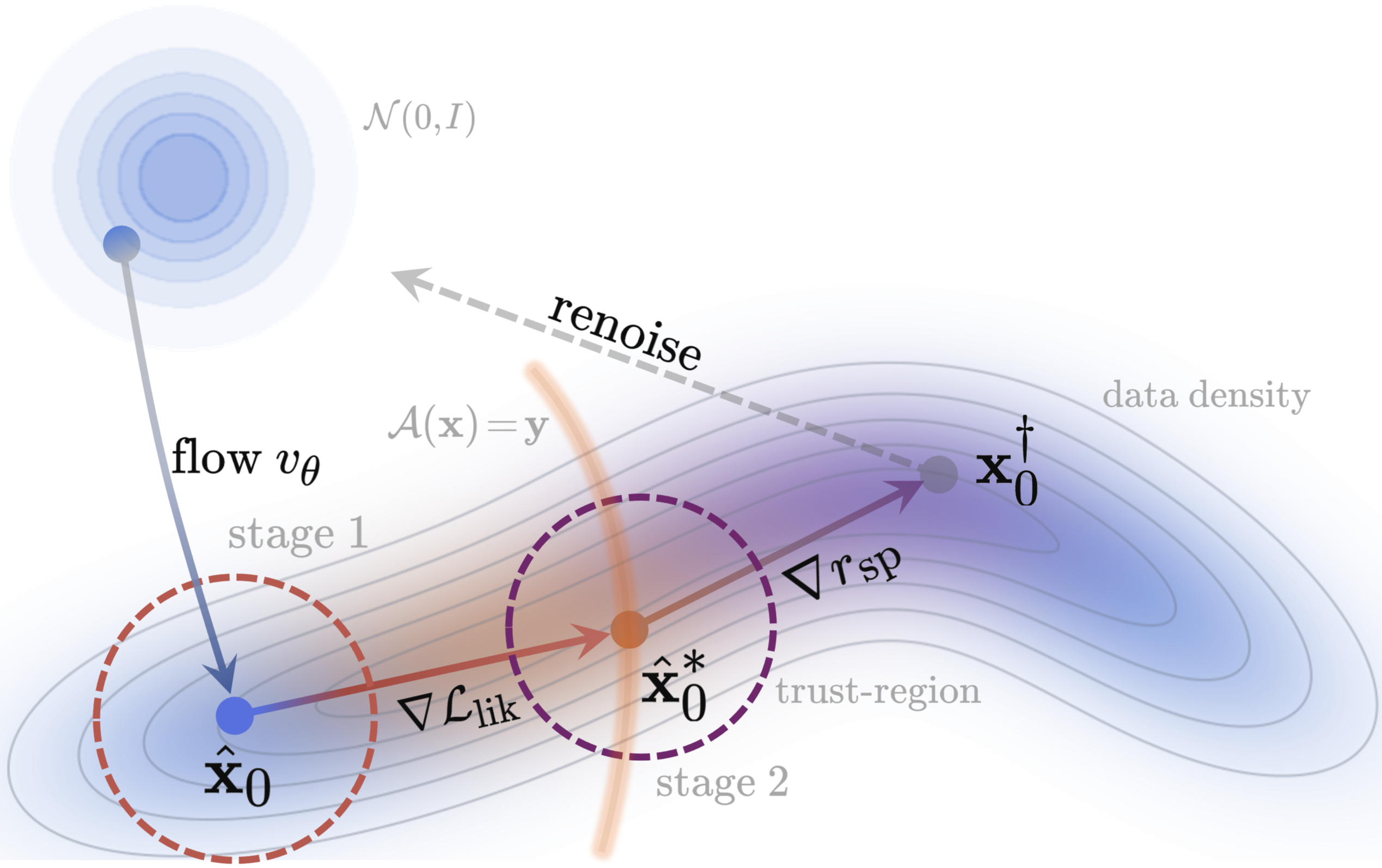}
    \vspace{-10pt}
    \caption{Decoupled refinement: the flow predicts $\hat\bx_0$, then likelihood and reward gradients, $\nabla \mathcal{L}_\mathrm{lik}$ and $\nabla r_\mathrm{sp}$, move the sample towards measurement fidelity (orange) and spatial consistency (purple).
    The optimized $\hat\bx_0^\dagger$ is then renoised and the process is repeated.}
    \label{fig:decoupled_refine}
\end{figure}
\subsection{Reward-tilting using Spatial Objectives}
\label{subsec:SpatialRewards}
Next, we seek to control perceptually relevant spatial attributes such as imaging width,
inter-channel coherence, and inter-channel level or phase differences.
We achieve this through a carefully designed spatial reward function $r_{\mathrm{sp}}(\bx_0)\in\mathbb{R}$ (discussed soon) which we use to \emph{tilt} the posterior in Eq. \ref{eq:daps-clean-posterior}. 
We first formulate this in a generalized setting as a KL-regularized reward-maximization problem, 
\begin{align}
\tilde p
=
\argmax_{q}\;
\mathbb{E}_{q}[r_{\mathrm{sp}}(\bx_0)]
-
\beta\,
D_{KL}\!\left(
q
\,\|\, 
p(\bx_0|\by,\bx_t)
\right)
\label{eq:variational}
\end{align}
where $\beta>0$ controls the trade-off between maximizing the reward and remaining close to the measurement-consistent posterior
$p(\bx_0|\bx_t,\by)$.
The solution $\tilde p$ to this variational problem takes the form of an exponential tilt
of the original posterior,
\begin{align}
\tilde p(\bx_0|\bx_t,\by)
\propto
p(\bx_0|\by,\bx_t)
\exp\left(
\frac{1}{\beta}r_{\mathrm{sp}}(\bx_0)
\right)
\label{eq:tilt}
\end{align}
Thus, the reward reweights the posterior toward stereo signals with the desired
spatial characteristics, while the KL penalty prevents the tilted distribution
from deviating arbitrarily from the measurement-consistent posterior in Eq. \ref{eq:daps-clean-posterior}.
Using the Gaussian factors, the negative log-density in Eq. \ref{eq:tilt} can be written as the following objective
\begin{align}
\mathcal{L}(\bx_0)
&=
\underbrace{\tfrac{1}{2\sigma_y^2}\|\oA(\bx_0)-\by\|^2}_{\text{data fidelity $= \nabla\mathcal{L}_\mathrm{lik}$}}
+
\underbrace{\tfrac{1}{2r_t^2}\|\bx_0-\hat{\bx}_0(\bx_t)\|^2}_{\text{prior proximity}}
\nonumber
-
\underbrace{\tfrac{1}{\beta}r_{\mathrm{sp}}(\bx_0)}_{\text{reward}}
\end{align}
We draw samples from $\tilde p$ and renoise the resulting clean signal to a lower noise level,
repeating until the noise level $t\rightarrow0$.

\subsection{Two-stage Decoupled Refinement}
\label{subsec:var_split}
The first two terms in $\mathcal{L}(\bx_0)$ promote measurement consistency and proximity to the flow's denoised prediction; the reward controls stereo spatial characteristics.
Initial experiments showed that joint optimization is unstable: the non-convex spatial rewards in Section \ref{sec:spatial_rewards} strongly interfere with measurement consistency before a plausible reconstruction is established.
We therefore introduce a \emph{decoupled refinement} strategy (Fig. \ref{fig:decoupled_refine}), first obtaining a measurement-consistent estimate regularized by the flow prior:
\begin{align}
\bx_0^\star
=
\argmin_{\bx_0}
\left\{
\tfrac{1}{2\sigma_y^2}
\|\oA(\bx_0)-\by\|_2^2
+
\tfrac{1}{2r_t^2}
\|\bx_0-\hat{\bx}_0(\bx_t)\|_2^2
\right\}.
\label{eq:stage1}
\end{align}
Once consistency has been established, we then seek a higher-reward stereo sample within a trust-region of this estimate:
\begin{align}
\max_{\bx_0}\; r_{\mathrm{sp}}(\bx_0)
\quad\text{s.t.}\quad
\|\bx_0-\bx_0^\star\|_2^2\leq\rho^2.
\label{eq:trust-region}
\end{align}
This constrains reward-driven modifications to remain within a neighborhood of the measurement-consistent estimate.
By performing a Lagrangian relaxation \cite{vandenberghe2004convex} of the trust-region constraint, we arrive at the proximal refinement objective
\begin{align}
\bx_0^\dagger
=
\argmin_{\bx_0}
\left\{
-\frac{1}{\beta}r_{\mathrm{sp}}(\bx_0)
+
\frac{\eta}{2}
\|\bx_0-\bx_0^\star\|_2^2
\right\},
\label{eq:stage2}
\end{align}
where $\eta>0$ penalizes departures from $\bx_0^\star$ and
$\beta$ scales the reward. 
This decoupled refinement provides a tractable surrogate for reward-tilted inference.
The resulting clean estimate $\bx_0^\dagger$ is re-noised to the next, lower noise level, and the procedure is repeated (Algorithm \ref{alg:rt_flow}).
Next, we describe the exact formulation of our spatial reward.

\begin{table*}[ht]
\centering
\caption{Quantitative results: MR-STFT and SI-SDR measure mono consistency; Spatial FD,
ILD and IC distance evaluate imaging quality. S3A and Parametric stereo are mono-preserving (\textbf{MP}) by design. Best shown in \textbf{bold} and second-best \underline{underlined}.}
\label{tab:main_results}

\renewcommand{\arraystretch}{0.95}
\setlength{\tabcolsep}{2.5pt}
\small

\begin{tabular}{lcccccccc}
\toprule
\textbf{Method} &
\textbf{Training-free} &
\textbf{MP} &
\textbf{MR-STFT} $\downarrow$ &
\textbf{SI-SDR (dB)} $\uparrow$ &
\textbf{SI-SDR Mid (dB)} $\uparrow$ &
\textbf{Spatial FD} $\downarrow$ &
\textbf{ILD Dist.} $\downarrow$ &
\textbf{IC Dist.} $\downarrow$ \\
\midrule

S3A (DSP)
& --
& \textcolor{green!50!black}{$\checkmark$}
& 0.290 & 1.44 & 24.14 & 93.9 & 2.203 & 0.246 \\
\midrule

\rowcolor{gray!10}
Parametric Stereo
& \textcolor{red}{$\times$}
& \textcolor{green!50!black}{$\checkmark$}
& 0.201 & 7.617 & 20.570 & 358.730 & 2.638 & 0.240 \\
\midrule

DiffStereo (DDIM)
& \textcolor{red}{$\times$}
& \textcolor{red}{$\times$}
& 0.581 & -2.16 & 1.79 & 53.3 & 1.460 & 0.343 \\

\rowcolor{gray!10}
DiffStereo (DDPM)
& \textcolor{red}{$\times$}
& \textcolor{red}{$\times$}
& 0.4550 & 0.569 & 4.34 & 43.16 & 1.552 & 0.263 \\
\midrule

FlowDPS
& \textcolor{green!50!black}{$\checkmark$}
& \textcolor{red}{$\times$}
& 0.657 & -18.036 & -17.045 & 25.659 & 1.36 & 0.25 \\
\midrule

\textbf{{\method} - Small (Ours)}
& \textcolor{green!50!black}{$\checkmark$}
& \textcolor{red}{$\times$}
& \underline{0.2223}
& \underline{4.6920}
& \textbf{8.34}
& \textbf{20.3344}
& \textbf{1.3744}
& \textbf{0.1693} \\

\rowcolor{gray!10}
\textbf{{\method} - Medium (Ours)}
& \textcolor{green!50!black}{$\checkmark$}
& \textcolor{red}{$\times$}
& \textbf{0.2088}
& \textbf{5.042}
& \underline{8.15}
& \underline{34.6924}
& \underline{1.543}
& \underline{0.233} \\

\bottomrule
\end{tabular}
\vspace{-10pt}
\end{table*}

\newcolumntype{Y}{>{\centering\arraybackslash}X}
\begin{table*}[t]
\centering
\caption{\text{Joint restoration and upmixing} from degraded mono observations. ``Degraded'' is the
observation (band-limited / clipped mono). Mid SI-SDR is
in-band ($\leq$cutoff) for low-pass, full-band for clipping. HF MR-STFT
($>$cutoff) is undefined for full-band clipping.}
\label{tab:restoration}
\renewcommand{\arraystretch}{0.95}
\setlength{\tabcolsep}{2pt}
\small
\begin{tabularx}{\textwidth}{@{}l l Y Y Y Y Y Y@{}}
\toprule
\textbf{Degradation} & \textbf{Method} &
\textbf{Mid In-band SI-SDR}$\uparrow$ &
\textbf{Side MR-STFT}$\downarrow$ &
\textbf{IC dist}$\downarrow$ &
\textbf{ILD dist}$\downarrow$ &
\textbf{Spatial FD}$\downarrow$ &
\textbf{HF MR-STFT}$\downarrow$ \\
\midrule
 & Degraded & 38.87 & 1.211 & 0.402 & 1.473 & 158.4 & 1.112 \\
\rowcolor{gray!10}
\cellcolor{white}\multirow{-2}{*}{LPF 8\,kHz} & {\method} & 9.36 & \textbf{0.653} & \textbf{0.149} & \textbf{1.332} & \textbf{18.6} & \textbf{0.744} \\
\midrule
 & Degraded & 40.02 & 1.211 & 0.402 & 1.473 & 158.4 & 1.045 \\
\rowcolor{gray!10}
\cellcolor{white}\multirow{-2}{*}{LPF 11\,kHz} & {\method} & 9.06 & \textbf{0.672} & \textbf{0.158} & \textbf{1.353} & \textbf{20.1} & \textbf{0.707} \\
\midrule
 & Degraded & 17.73 & 1.211 & 0.402 & 1.473 & 158.4 & -- \\
\rowcolor{gray!10}
\cellcolor{white}\multirow{-2}{*}{Clip} & {\method} & 6.74 & \textbf{0.655} & \textbf{0.157} & \textbf{1.364} & \textbf{18.6} & -- \\
\bottomrule
\end{tabularx}
\end{table*}

\subsection{Designing Spatial Rewards}
\label{sec:spatial_rewards}

Our \emph{spatial reward} $r_{\mathrm{sp}}$ compares the inter-channel statistics of a candidate $\hat{\bx}_0$ with those of a reference
$\bx^{\mathrm{ref}}$, from which target reward parameters are estimated. 
Two principles guide its design:
(1) cues are evaluated within perceptual frequency bands to capture frequency-dependent localization and width; (2) each cue is represented by its distribution over time, capturing fluctuations that a single across-time average would miss.

\textbf{Band-wise Spatial Covariance.}
Let $\bX_c\in\mathbb{C}^{K\times N}$ denote the STFT of channel $c\in{L,R}$, where $X_c[k,n]$ denotes frequency bin $k$ and time frame $n$. We collect the two channels as
$\bX[k,n]=[X_L[k,n],X_R[k,n]]^{\top}\in\mathbb{C}^{2}$.
Let $\bW\in\mathbb{R}^{N_b\times K}$ denote an ERB-spaced \cite{erb} banding matrix, where $W_{b,k}$ specifies the contribution of frequency bin $k$ to band $b$. For each band and time frame, we compute the spatial covariance matrix (SCM) $\bR_{b,n}$
with diagonal entries $E^{LL}_{b,n}$ and $E^{RR}_{b,n}$ and off-diagonal entry $C_{b,n}$,
{\small
\begin{align}
\bR_{b,n}
=
\sum_k W_{b,k}\bX[k,n]\bX[k,n]^{\mathsf H}
=
\begin{bmatrix}
E^{LL}_{b,n} & C_{b,n}\\
C_{b,n}^{*} & E^{RR}_{b,n}
\end{bmatrix}
\label{eq:scm}
\end{align}
}
\noindent Each $\bR_{b,n}\in\mathbb{C}^{2\times2}$ satisfies $\bR_{b,n}=\bR_{b,n}^{\mathsf H}\succeq 0$ and summarizes the inter-channel energy and correlation structure within a perceptual frequency band. 
The SCM is differentiable with respect to the waveform, enabling gradient-based refinement \cite{adam}.

\textbf{Fitting Spatial Cue Distributions.}
From each SCM, we extract two complementary spatial cues
\cite{breebaart2005parametric}: inter-channel coherence (IC), given by the
normalized off-diagonal magnitude, and inter-channel level difference (ILD),
given by the ratio of channel energies:

{\footnotesize
\begin{align}
\rho_{b,n}
&=
\frac{|C_{b,n}|}
{\sqrt{E^{LL}_{b,n}E^{RR}_{b,n}}+\epsilon},
\quad
\Delta_{b,n}
=
10\log_{10}
\frac{E^{LL}_{b,n}+\epsilon}
{E^{RR}_{b,n}+\epsilon}
\label{eq:iid_iic}
\end{align}}

Low coherence indicates greater inter-channel decorrelation and more a diffuse spatial image, whereas ILD captures frequency-dependent
left--right panning.
Rather than averaging these cues over time, we model $\{\rho_{b,n}\}_{n=1}^{N}$ and $\{\Delta_{b,n}\}_{n=1}^{N}$ as samples from
per-band distributions. 
Since coherence is bounded on $[0,1]$, we model it as a Beta distribution \cite{johnson1995continuous},
$\rho_{b,\cdot}
\sim
\mathrm{Beta}(\alpha_b,\beta_b).$
The parameters are estimated by moment matching from the empirical mean $\hat{\mu}_b$ and variance $\hat{v}_b$: 
$\nu_b
=
\frac{
\hat{\mu}_b(1-\hat{\mu}_b)
}{
\hat{v}_b
}
-1,\,
\alpha_b
=
\hat{\mu}_b\nu_b, \,
\beta_b
=
(1-\hat{\mu}_b)\nu_b$.
ILD is unbounded and approximately symmetric in the log-amplitude domain, and is modeled as a Gaussian, $\Delta_{b,\cdot} \sim \mathcal{N}(\mu_b,\sigma_b^2)$: 
$\mu_b
=
\frac{1}{N}
\sum_{n=1}^{N}\Delta_{b,n},
\sigma_b^2
=
\hat{\mathrm{Var}}_{n}
\left[
\Delta_{b,n}
\right].$
Applying the same estimator to the reference spectrogram $\bX^{\mathrm{ref}}$ yields target parameters
$(\alpha_b^\star,\beta_b^\star,\mu_b^\star,\sigma_b^\star)$ for each
frequency band. 
Thus, the spatial target is specified entirely by a single
reference excerpt and requires no paired supervision. 
Additional control on the stereo width can be achieved by modifying the target coherence before
fitting the Beta parameters. 
Specifically, a steering coefficient $\kappa$ adjusts the target mean coherence as
$
\rho_b^\star
\leftarrow
1-
\kappa
\left(
1-\rho_b^\star
\right),
\label{eq:coherence-steering}
$
with the resulting value constrained to $[0,1]$ before refitting the target distribution.

\textbf{Spatial Reward.}
Let $(\hat{\alpha}_b,\hat{\beta}_b,\hat{\mu}_b,\hat{\sigma}_b)$ denote the
parameters estimated from a candidate clean stereo signal $\bx_0$, and
let $(\alpha_b^\star,\beta_b^\star,\mu_b^\star,\sigma_b^\star)$ denote the
corresponding target parameters. We define the spatial reward as the negative
average KL divergence between the candidate and target cue distributions,
{\footnotesize
\begin{align}
r_{\mathrm{sp}}(\bx_0)
&=
-\frac{1}{N_b}
\sum_{b=1}^{N_b}
\Bigg[
D_{\mathrm{KL}}
\Big(
\mathrm{Beta}(\hat{\alpha}_b,\hat{\beta}_b)
\,\|\,
\mathrm{Beta}(\alpha_b^\star,\beta_b^\star)
\Big)
\nonumber\\
&\qquad\qquad
+
\lambda\,
D_{\mathrm{KL}}
\Big(
\mathcal{N}(\hat{\mu}_b,\hat{\sigma}_b^2)
\,\|\,
\mathcal{N}(\mu_b^\star,\sigma_b^{\star 2})
\Big)
\Bigg]
\label{eq:spatial_reward}
\end{align}}
where $\lambda$ balances the contributions of the coherence and level cues.
The complete mapping from the candidate waveform $\hat{\bx}_0$ to
the cue-distribution parameters is differentiable, $\nabla_{\hat{\bx}_0}r_{\mathrm{sp}}$ can be obtained by backpropagation and
used directly for reward-based refinement (see App. \ref{app:KL_impl}). 
The distribution means determine the average coherence and left--right balance, while the variances and
Beta shape parameters capture temporal fluctuations to mimic realistic stereo images.

\section{Experiments}
\label{sec:Experiments}
\noindent \textbf{Flow Prior.}
We use Stable Audio 3~\cite{evans2026stableaudio3} (medium and small) as a latent stereo prior at $44.1$\,kHz, with classifier-free guidance
disabled \cite{cfg}. 
We retain its native $8$-step schedule, applying $15$ measurement-consistency and $5$ reward-refinement iterations per step (App. \ref{app:perf_impr}).

\noindent \textbf{Datasets.}
We evaluate on real-world stereo recordings from the MUSDB18-HQ \cite{musdb18-hq} test set ($44.1$\,kHz), remixed without vocals and peak-normalized to $-0.5$\,dBFS. 
We use $200$ disjoint $5$\,s excerpts that have integrated loudness (ITU-R BS.1770) above $-30$\,LUFS and within $8$\,LU of the parent track, $\geq95\%$ of $100$\,ms frames above a track-relative silence floor, side-to-mid energy above $-25$\,dB, crest factor
$6$--$22$\,dB, spectral flatness $<0.6$, and $\geq2\%$ energy above $11$\,kHz. 
We select 4 excerpts per test track with $\geq10$\,s separation.

\noindent \textbf{Tasks.}
We consider two distinct settings: 1) \emph{mono-to-stereo upmixing}, 2) \emph{joint upmixing and restoration} where $\mathcal{A}$ additionally applies a degradation $\oD$, an FIR low-pass filter at $8$ and $11$\,kHz (App. \ref{app:lpf}), or soft clipping, before the downmix $\oM$.
Bandwidth-extension must therefore be performed alongside recovering the stereo image.
The same prior and guidance are used but $\mathcal{A}$ changes, with no retraining.

\noindent \textbf{Metrics.} The generated stereo audio were compared with their clean stereo references using complementary objective metrics. 
MR-STFT \cite{mrstft, auraloss} measures spectral reconstruction error, while SI-SDR (overall and mid-channel) \cite{le2019sdr} evaluates waveform fidelity independently of signal scale. Stereo-image accuracy was assessed using ILD error and IC error, and Spatial Fréchet Distance, which compares the joint distribution of ERB-band ILD and IC features.

\noindent \textbf{Baselines.}
Our baselines include: 
S3A~\cite{schlecht2018optimized}, a classical mono-preserving decorrelator.
DiffStereo~\cite{Zhang2025DiffStereoEM},  a diffusion-based mono-to-stereo upmixer;
we report both DDIM~\cite{ddim} and DDPM~\cite{ddpm} sampling.
We also include a parametric upmixer that predicts stereo parameters~\cite{serra}.
Finally, we compare against FlowDPS \cite{kim2025flowdps}, a one-stage generative inverse solver.

\subsection{Results: Objective and Subjective}
\label{sec:Results}

\noindent\textbf{Mono-to-stereo upmixing.}
In Table~\ref{tab:main_results}, {\method}\textbf{-Small} achieves the
lowest Spatial FD (20.33) and IC distance (0.169), with better
waveform fidelity than DiffStereo and FlowDPS. S3A and
Parametric Stereo are designed to preserve the mono component but yield poorer spatial metrics, while FlowDPS combines competitive spatial statistics but with extreme waveform distortion. 
{\method}\textbf{-Medium} improves MR-STFT and overall SI-SDR slightly but underperforms Small on all spatial metrics, indicating that greater model size does not uniformly improve upmixing. 
FlowDPS achieves a marginally lower ILD distance than Small (1.360 versus 1.374).

\noindent\textbf{Joint restoration and upmixing.}
Across bandwidth-limited and clipped inputs (Table~\ref{tab:restoration}), {\method} reduces Spatial FD
from 158.4 to 18.6--20.1 and side MR-STFT from 1.211 to 0.653--0.672. 
For 8 and 11\,kHz low-pass inputs, HF MR-STFT improves from 1.112 to 0.744 and 1.045 to 0.707, respectively.
These gains accompany reduced in-band mid SI-SDR; for clipping, full-band mid SI-SDR falls from 17.73 to 6.74\,dB. 
The results demonstrate improved spatial and high-frequency reconstruction, with an expected loss in mid-signal fidelity due to our generative approach.

\begin{figure}[ht]
    \centering
    \includegraphics[width=0.99\columnwidth]{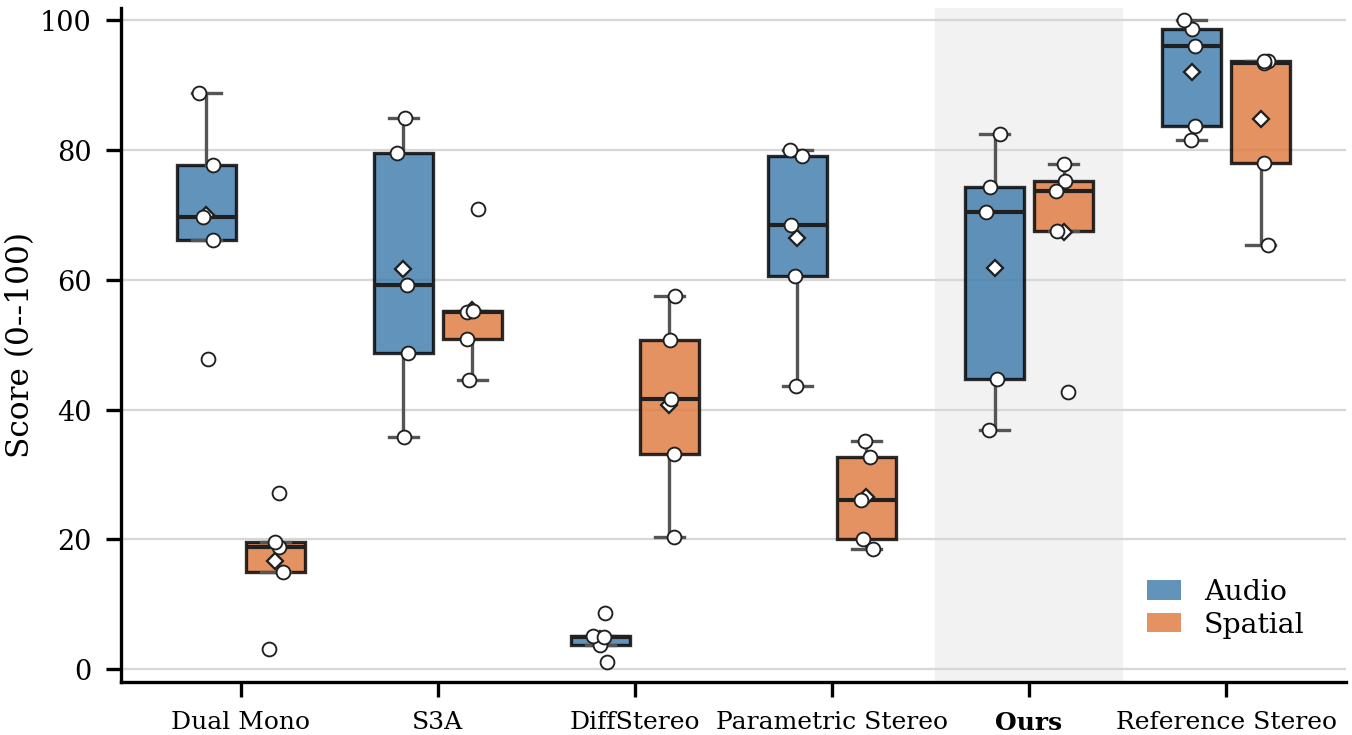}
    \vspace{-10pt}
    \caption{Subjective tests across baselines: Boxes show the interquartile range, lines denote medians, and diamonds indicate means}
    \label{fig:boxplot}
\end{figure}

\noindent\textbf{Listening Tests.} We conducted a double-blind test in which 5 participants evaluated 6 versions of 8 music excerpts: the hidden stereo reference,
{\method}, S3A, DiffStereo, Parametric Stereo, and a dual-mono anchor.
Labels were anonymized and presentation order randomized for each
participant.
Listeners independently rated audio quality (noise, distortion, coloration) and spatial imaging (width, placement, balance, and stability) on continuous 0--100 scales.
Fig. \ref{fig:boxplot} summarizes the screened ratings. 
The extremes behave as expected: the hidden reference scores highest on both axes ($92.0$, $84.9$), while the dual-mono attains the best audio
quality of any non-reference system ($70.0$) but the weakest spatial
imaging ($16.7$).
DiffStereo collapses in audio quality ($4.7$), indicating that its widening
comes with severe artifacts. 
{\method} achieves the highest spatial rating among all upmixing systems ($67.4$ vs. $55.3$ for S3A) at a competitive audio quality ($61.8$) relative to baselines.

\vspace{-1pt}
\section{Discussion}
\label{sec:conclusion}

\noindent\textbf{Conclusion.}
We presented a training-free framework for mono-to-stereo upmixing
that combines a pretrained flow prior with inference-time spatial
rewards. 
By decoupling measurement-consistent reconstruction from
spatial refinement, the method supports controllable stereo
generation without task-specific retraining. 
A composite forward operator extends the same framework to joint restoration and
upmixing. 
Experiments and listening tests on real-world stereo recordings from MUSDB18-HQ demonstrate improved spatial
imaging and competitive audio quality across tasks.

\noindent \textbf{Follow-up research.}
To develop this generative upmixing framework, we assume a known forward operator $\oA$ at inference-time. 
We aim to relax these assumptions to enable blind upmixing using contrastive surrogate likelihoods that have shown promise in such challenging settings \cite{coguide,audiocoguide}. 
Further, our method is not limited to mono-to-stereo upmixing but extends naturally to multichannel audio generation when high-fidelity surround priors (e.g., 5.1, 7.1) become available. 
We believe {\method} could initiate conversations along many such directions.

\vfill\pagebreak

\clearpage
{\footnotesize
\bibliographystyle{IEEEbib}
\bibliography{refs}}

\begin{thebibliography}{10}

\bibitem{kendall1995decorrelation}
Gary~S. Kendall,
\newblock ``The decorrelation of audio signals and its impact on spatial imagery,''
\newblock {\em Computer Music Journal}, vol. 19, no. 4, pp. 71--87, 1995.

\bibitem{fink2015downmix}
Marco Fink, Sebastian Kraft, and Udo Z{\"o}lzer,
\newblock ``Downmix-compatible conversion from mono to stereo in time- and frequency-domain,''
\newblock in {\em Proceedings of the 18th International Conference on Digital Audio Effects (DAFx-15)}, Trondheim, Norway, 2015, pp. 411--414.

\bibitem{schlecht2018optimized}
Sebastian~Jiro Schlecht, Benoit Alary, Vesa V{\"a}lim{\"a}ki, and Emanu{\"e}l A.~P. Habets,
\newblock ``Optimized velvet-noise decorrelator,''
\newblock in {\em Proceedings of the 21st International Conference on Digital Audio Effects (DAFx-18)}, Aveiro, Portugal, 2018, pp. 87--94.

\bibitem{alary2017velvet}
Benoit Alary, Archontis Politis, and Vesa V{\"a}lim{\"a}ki,
\newblock ``Velvet-noise decorrelator,''
\newblock in {\em Proceedings of the 20th International Conference on Digital Audio Effects (DAFx-17)}, Edinburgh, United Kingdom, 2017, pp. 405--411.

\bibitem{he2015ambient}
Jianjun He, Woon-Seng Gan, and Ee-Leng Tan,
\newblock ``Primary-ambient extraction using ambient spectrum estimation for immersive spatial audio reproduction,''
\newblock {\em IEEE/ACM Transactions on Audio, Speech, and Language Processing}, vol. 23, no. 9, pp. 1431--1444, Sept. 2015.

\bibitem{breebart}
Christof Faller and Jeroen Breebaart,
\newblock ``Binaural reproduction of stereo signals using upmixing and diffuse rendering,''
\newblock 10 2011, vol.~2.

\bibitem{Zhang2025DiffStereoEM}
Suqi Zhang, Zheqi Dai, Yongyi Zang, Yin Cao, and Qiuqiang Kong,
\newblock ``Diffstereo: End-to-end mono-to-stereo audio generation with diffusion transformer,''
\newblock {\em Interspeech 2025}, 2025.

\bibitem{ho2020denoising}
Jonathan Ho, Ajay Jain, and Pieter Abbeel,
\newblock ``Denoising diffusion probabilistic models,''
\newblock {\em Advances in neural information processing systems}, vol. 33, pp. 6840--6851, 2020.

\bibitem{hu2021lora}
Edward~J Hu, Yelong Shen, Phillip Wallis, Zeyuan Allen-Zhu, Yuanzhi Li, Shean Wang, Lu~Wang, and Weizhu Chen,
\newblock ``Lora: Low-rank adaptation of large language models,''
\newblock {\em arXiv preprint arXiv:2106.09685}, 2021.

\bibitem{serra}
Joan Serr{\`a}, Davide Scaini, Santiago Pascual, Daniel Arteaga, Jordi Pons, Jeroen Breebaart, and Giulio Cengarle,
\newblock ``Mono-to-stereo through parametric stereo generation,''
\newblock {\em arXiv preprint arXiv:2306.14647}, 2023.

\bibitem{hadamard}
Jacques {Hadamard} and Philip~M. {Morse},
\newblock ``{Lectures on Cauchy's Problem in Linear Partial Differential Equations},''
\newblock {\em Physics Today}, vol. 6, no. 8, pp. 18, Jan. 1953.

\bibitem{dps}
Hyungjin Chung, Jeongsol Kim, Michael~T Mccann, Marc~L Klasky, and Jong~Chul Ye,
\newblock ``Diffusion posterior sampling for general noisy inverse problems,''
\newblock {\em arXiv preprint arXiv:2209.14687}, 2022.

\bibitem{musdb18-hq}
Zafar Rafii, Antoine Liutkus, Fabian-Robert Stöter, Stylianos~Ioannis Mimilakis, and Rachel Bittner,
\newblock ``Musdb18-hq - an uncompressed version of musdb18,'' Aug. 2019.

\bibitem{ldm}
Robin Rombach, Andreas Blattmann, Dominik Lorenz, Patrick Esser, and Bj{\"o}rn Ommer,
\newblock ``High-resolution image synthesis with latent diffusion models,''
\newblock in {\em Proceedings of the IEEE/CVF conference on computer vision and pattern recognition}, 2022, pp. 10684--10695.

\bibitem{EnglHankeNeubauer1996}
Heinz~W. Engl, Martin Hanke, and Andreas Neubauer,
\newblock {\em Regularization of Inverse Problems}, vol. 375 of {\em Mathematics and Its Applications},
\newblock Springer, Dordrecht, 1996.

\bibitem{lipman2022flow}
Yaron Lipman, Ricky~TQ Chen, Heli Ben-Hamu, Maximilian Nickel, and Matt Le,
\newblock ``Flow matching for generative modeling,''
\newblock {\em arXiv preprint arXiv:2210.02747}, 2022.

\bibitem{liu2022flow}
Xingchao Liu, Chengyue Gong, and Qiang Liu,
\newblock ``Flow straight and fast: Learning to generate and transfer data with rectified flow,''
\newblock {\em arXiv preprint arXiv:2209.03003}, 2022.

\bibitem{Batchelor_2000}
G.~K. Batchelor,
\newblock {\em An Introduction to Fluid Dynamics},
\newblock Cambridge Mathematical Library. Cambridge University Press, 2000.

\bibitem{albergo2025stochastic}
Michael Albergo, Nicholas~M Boffi, and Eric Vanden-Eijnden,
\newblock ``Stochastic interpolants: A unifying framework for flows and diffusions,''
\newblock {\em Journal of Machine Learning Research}, vol. 26, no. 209, pp. 1--80, 2025.

\bibitem{score}
Yang Song, Jascha Sohl-Dickstein, Diederik~P Kingma, Abhishek Kumar, Stefano Ermon, and Ben Poole,
\newblock ``Score-based generative modeling through stochastic differential equations,''
\newblock {\em arXiv preprint arXiv:2011.13456}, 2020.

\bibitem{kim2025flowdps}
Jeongsol Kim, Bryan~Sangwoo Kim, and Jong~Chul Ye,
\newblock ``Flowdps: Flow-driven posterior sampling for inverse problems,''
\newblock in {\em 2025 IEEE/CVF International Conference on Computer Vision (ICCV)}. IEEE, 2025, pp. 12328--12337.

\bibitem{ddpm}
Jonathan Ho, Ajay Jain, and Pieter Abbeel,
\newblock ``Denoising diffusion probabilistic models,''
\newblock {\em Advances in neural information processing systems}, vol. 33, pp. 6840--6851, 2020.

\bibitem{inversebench}
Hongkai Zheng, Wenda Chu, Bingliang Zhang, Zihui Wu, Austin Wang, Berthy~T. Feng, Caifeng Zou, Yu~Sun, Nikola Kovachki, Zachary~E. Ross, Katherine~L. Bouman, and Yisong Yue,
\newblock ``Inversebench: Benchmarking plug-and-play diffusion priors for inverse problems in physical sciences,'' 2025.

\bibitem{daps}
Bingliang Zhang, Wenda Chu, Julius Berner, Chenlin Meng, Anima Anandkumar, and Yang Song,
\newblock ``Improving diffusion inverse problem solving with decoupled noise annealing,''
\newblock in {\em 2025 IEEE/CVF Conference on Computer Vision and Pattern Recognition (CVPR)}. IEEE, 2025, pp. 20895--20905.

\bibitem{song2019generative}
Yang Song and Stefano Ermon,
\newblock ``Generative modeling by estimating gradients of the data distribution,''
\newblock {\em Advances in neural information processing systems}, vol. 32, 2019.

\bibitem{efron2011tweedie}
Bradley Efron,
\newblock ``Tweedie's formula and selection bias,''
\newblock {\em Journal of the American Statistical Association}, vol. 106, no. 496, pp. 1602--1614, 2011.

\bibitem{vandenberghe2004convex}
Lieven Vandenberghe,
\newblock {\em Convex optimization}, vol.~1.

\bibitem{erb}
Stephen Neely,
\newblock ``Signals, sound, and sensation, by william m. hartmann,''
\newblock {\em Physics Today - PHYS TODAY}, vol. 51, 01 1998.

\bibitem{adam}
Diederik~P. Kingma and Jimmy Ba,
\newblock ``Adam: A method for stochastic optimization,'' 2017.

\bibitem{breebaart2005parametric}
Jeroen Breebaart, Steven van~de Par, Armin Kohlrausch, and Erik Schuijers,
\newblock ``Parametric coding of stereo audio,''
\newblock {\em EURASIP Journal on Applied Signal Processing}, vol. 2005, pp. 1305--1322, 2005.

\bibitem{johnson1995continuous}
Norman~L. Johnson, Samuel Kotz, and Narayanaswamy Balakrishnan,
\newblock {\em Continuous Univariate Distributions}, vol.~2,
\newblock Wiley, 2 edition, 1995.

\bibitem{evans2026stableaudio3}
Zach Evans, Julian~D. Parker, Matthew Rice, CJ~Carr, Zack Zukowski, Josiah Taylor, and Jordi Pons,
\newblock ``Stable audio 3,'' 2026.

\bibitem{cfg}
Prafulla Dhariwal and Alexander Nichol,
\newblock ``Diffusion models beat gans on image synthesis,''
\newblock {\em Advances in neural information processing systems}, vol. 34, pp. 8780--8794, 2021.

\bibitem{mrstft}
Ryuichi Yamamoto, Eunwoo Song, and Jae-Min Kim,
\newblock ``Parallel wavegan: A fast waveform generation model based on generative adversarial networks with multi-resolution spectrogram,''
\newblock in {\em ICASSP 2020-2020 IEEE International Conference on Acoustics, Speech and Signal Processing (ICASSP)}. IEEE, 2020, pp. 6199--6203.

\bibitem{auraloss}
C.~Steinmetz and Joshua~D. Reiss,
\newblock ``auraloss: Audio-focused loss functions in pytorch,''
\newblock .

\bibitem{le2019sdr}
Jonathan Le~Roux, Scott Wisdom, Hakan Erdogan, and John~R Hershey,
\newblock ``Sdr--half-baked or well done?,''
\newblock in {\em ICASSP 2019-2019 IEEE International Conference on Acoustics, Speech and Signal Processing (ICASSP)}. IEEE, 2019, pp. 626--630.

\bibitem{ddim}
Jiaming Song, Chenlin Meng, and Stefano Ermon,
\newblock ``Denoising diffusion implicit models,''
\newblock {\em arXiv preprint arXiv:2010.02502}, 2020.

\bibitem{coguide}
Sattwik Basu, Chaitanya Amballa, Zhongweiyang Xu, Jorge Van{\v{c}}o~Sampedro, Srihari Nelakuditi, and Romit Roy~Choudhury,
\newblock ``Contrastive diffusion guidance for spatial inverse problems,''
\newblock in {\em International Conference on Learning Representations}, 2026, vol. 2026, pp. 85262--85290.

\bibitem{audiocoguide}
Sattwik Basu, Chaitanya Amballa, Zhongweiyang Xu, Jorge~Vanco Sampedro, Srihari Nelakuditi, and Romit~Roy Choudhury,
\newblock ``Blind audio restoration using contrastive diffusion guidance,''
\newblock in {\em ICML 2026 Workshop on Structured Probabilistic Inference and Generative Modeling (SPIGM)}, 2026.

\end{thebibliography}

\section{Appendix}

\subsection{Lowpass Filter Design}
\label{app:lpf}
In the joint upmixing and bandwidth restorstion task, $\mathcal{A}$ is a linear-phase FIR low-pass filter (Kaiser-windowed sinc,
80\,dB stopband attenuation, 200\,Hz transition band, $N=1107$ taps at
44.1\,kHz) with its $-6$\,dB point at $f_c$. It is applied by zero-padded
FFT convolution and compensated for its $(N-1)/2$-sample group delay, so it is
time-aligned, differentiable, and self-adjoint ($\mathcal{A}^{T}=\mathcal{A}$).

\subsection{Description of the Algorithm}
\label{app:algorithm}

We implement {\method} with its two-stage \emph{decoupled-refinement} algorithm as follows. 

\begin{algorithm}[ht]
\caption{Audio Upmixing using Reward-tilted Flows}
\label{alg:rt_flow}
\begin{algorithmic}

\Require Measurement $\by$, Operator $\oA$, Pre-trained Flow Model $\bv_\theta$, Spatial-reward $r_{\mathrm{sp}}$,
$\{t_i\}_{i=0}^{T},K_c,K_r,\beta,\eta$

\State $\bx_{t_T}\sim\mathcal{N}(\boldsymbol{0},\bI)$

\For{$i=T,T-1,\ldots,1$}

    \State \textbf{Flow denoising:}
    
    $\hat{\bx}_0
    \gets
    \bx_{t_i}-t_i\bv_\theta(\bx_{t_i},t_i)$

    \Statex
    \hspace{\algorithmicindent}
    \begin{minipage}{\dimexpr\linewidth-\algorithmicindent\relax}
    \begin{stagebox}{stageblue}

    \textbf{Stage 1: Measurement consistency} 
    ($K_c$ steps) 
    
    $\bx_0\gets\hat{\bx}_0$\par
    $\displaystyle
    \bx_0\gets
    \bx_0-\alpha_c\nabla_{\bx_0}
    \left[
    \frac{\|\oA(\bx_0)-\by\|_2^2}{2\sigma_y^2}
    +
    \frac{\|\bx_0-\hat{\bx}_0\|_2^2}{2r_{t_i}^2}
    \right]$\par
    $\bx_0^\star\gets\bx_0$

    \end{stagebox}
    \end{minipage}

    \Statex
    \hspace{\algorithmicindent}
    \begin{minipage}{\dimexpr\linewidth-\algorithmicindent\relax}
    \begin{stagebox}{stageorange}

    \textbf{Stage 2: Reward refinement}
    ($K_r$ steps)
    
    $\bx_0\gets\bx_0^\star$\par
    $\displaystyle
    \bx_0\gets
    \bx_0-\alpha_r\nabla_{\bx_0}
    \left[
    -\frac{1}{\beta}r_{\mathrm{sp}}(\bx_0)
    +
    \frac{\eta}{2}\|\bx_0-\bx_0^\star\|_2^2
    \right]$\par
    $\bx_0^\dagger\gets\bx_0$

    \end{stagebox}
    \end{minipage}

    \Statex
    \hspace{\algorithmicindent}
    \begin{minipage}{\dimexpr\linewidth-\algorithmicindent\relax}
    \begin{stagebox}{stagepurple}

    \textbf{Renoising:}
    
    $\beps\sim\mathcal{N}(\boldsymbol{0},\bI)$,\;
    $\bx_{t_{i-1}}
    \gets
    (1-t_{i-1})\bx_0^\dagger+t_{i-1}\beps$

    \end{stagebox}
    \end{minipage}

\EndFor

\State \Return $\bx_0^\dagger$

\end{algorithmic}
\end{algorithm}
\noindent Our algorithm operates in the latent space of the Stable Audio 3 model. 
For a regular (non-latent) diffusion or flow model, the decoder $\varphi$ is simply an identity function. 
The gradients are obtained by backpropagating through $\varphi$ and $\oA$.
We update the latent vectors using Adam. 
Adam normalizes each latent coordinate by using a smoothed
gradient magnitude. 
A single learning rate $\eta$ per optimization phase therefore works
whether a loss value is large (early i.e., far from the target) or small (during later stages), and whether its scale is set by the spatial reward or by the likelihood.

\subsection{Closed-form Expression for the Spatial Reward}
\label{app:KL_impl}

The reward described in Sec. \ref{sec:spatial_rewards} compares, in every ERB band $b$, the distribution of a cue in the
generated stereo with its distribution in the target through a KL-divergence.
Two things make this hard to optimize with gradients.
(i) We never observe $\hat{p}$ directly. The decoded audio only gives $N$ per-frame
samples $\{\rho_{b,n},\Delta_{b,n}\}_{n=1}^{N}$ ($N\approx216$ for a 5\,s chunk).
The KL between two sets of samples is undefined, and a histogram estimate is
piecewise constant in the audio, so its gradient is zero almost everywhere.
(ii) Even with a density for $\hat{p}$, the KL is still an integral. Estimating
it frame by frame evaluates the Beta log-density $\log(1-\rho_{b,n})$, which
diverges as $\rho\to1$, i.e., exactly in the near-mono frequency bands we want to widen.

\noindent \textbf{Proposed solution.}
We summarize each band by a few smooth statistics, fit a parametric
distribution to them by moment matching, and compare the fitted candidate
with the fitted target through a closed-form KL. 
We derive the integral and ensure the sampler only evaluates the final resulting expression. 
The whole transformation sequence from latent to loss is then a composition of differentiable functions, and thus
autograd is able to produce $\nabla_{\bx_0}r_{\mathrm{sp}}$.

\noindent \textbf{Closed-form KL for the ILD.}
The candidate and target ILD densities are
$\hat{p}(\Delta)=\N(\Delta;\hat{\mu}_b,\hat{\sigma}_b^2)$ and
$p^{\star}(\Delta)=\N(\Delta;\mu_b^{\star},\sigma_b^{\star2})$, with
$\N(\Delta;\mu,\sigma^2)=(2\pi\sigma^2)^{-1/2}\exp\!\big(-(\Delta-\mu)^2/(2\sigma^2)\big)$.
The $2\pi$ factors cancel in the log-ratio:
\begin{align}
\log\frac{\hat{p}(\Delta)}{p^{\star}(\Delta)}
&=
\log\frac{\sigma_b^{\star}}{\hat{\sigma}_b}
-\frac{(\Delta-\hat{\mu}_b)^2}{2\hat{\sigma}_b^2}
+\frac{(\Delta-\mu_b^{\star})^2}{2\sigma_b^{\star2}}.
\nonumber
\end{align}
The KL is the expectation of this under $\hat{p}$:
\begin{align}
D^{\Delta}_b
&=
\mathbb{E}_{\hat{p}}\!\left[\log\frac{\hat{p}(\Delta)}{p^{\star}(\Delta)}\right]
\nonumber\\
&=
\log\frac{\sigma_b^{\star}}{\hat{\sigma}_b}
-\frac{\mathbb{E}_{\hat{p}}[(\Delta-\hat{\mu}_b)^2]}{2\hat{\sigma}_b^2}
+\frac{\mathbb{E}_{\hat{p}}[(\Delta-\mu_b^{\star})^2]}{2\sigma_b^{\star2}}.
\nonumber
\end{align}
The first expectation is the variance of $\hat{p}$, i.e.\ $\hat{\sigma}_b^2$.
For the second, we first add and subtract $\hat{\mu}_b$ and use
$\mathbb{E}_{\hat{p}}[\Delta-\hat{\mu}_b]=0$:
\begin{align}
\mathbb{E}_{\hat{p}}[(\Delta-\mu_b^{\star})^2]
&=
\mathbb{E}_{\hat{p}}\big[\big((\Delta-\hat{\mu}_b)+(\hat{\mu}_b-\mu_b^{\star})\big)^2\big]
\nonumber\\
&=
\hat{\sigma}_b^2
+2(\hat{\mu}_b-\mu_b^{\star})\,\underbrace{\mathbb{E}_{\hat{p}}[\Delta-\hat{\mu}_b]}_{=0}
\nonumber\\
&\quad
+(\hat{\mu}_b-\mu_b^{\star})^2
\nonumber\\
&=
\hat{\sigma}_b^2+(\hat{\mu}_b-\mu_b^{\star})^2.
\nonumber
\end{align}
Substituting both expectations,
\begin{align}
D^{\Delta}_b
&=
\log\frac{\sigma_b^{\star}}{\hat{\sigma}_b}
-\frac{1}{2}
+\frac{\hat{\sigma}_b^{2}+(\hat{\mu}_b-\mu_b^{\star})^{2}}
     {2\sigma_b^{\star 2}}.
\label{eq:gauss_kl}
\end{align}
The term $(\hat{\mu}_b-\mu_b^{\star})^2$ penalizes a wrong mean panning, and
the $\hat{\sigma}_b$ terms penalize a wrong amount of ILD fluctuation.

\noindent \textbf{Closed-form KL for the IC.}
Simplifying the KL-divergence in IC reward is somewhat more involved. 
Here, the Beta density and the Beta function are
\begin{align}
p(\rho;\alpha,\beta)
&=\frac{\rho^{\alpha-1}(1-\rho)^{\beta-1}}{\mathcal{B}(\alpha,\beta)},
\quad \rho\in(0,1),
\nonumber\\
\mathcal{B}(\alpha,\beta)
&=\int_0^1\rho^{\alpha-1}(1-\rho)^{\beta-1}\,d\rho
=\frac{\Gamma(\alpha)\Gamma(\beta)}{\Gamma(\alpha+\beta)},
\label{eq:beta_fn}
\end{align}
where $\Gamma(\cdot)$ is the Gamma function. Taking the logarithm of the density,
\begin{align}
\log p(\rho;\alpha,\beta)
&=(\alpha-1)\log\rho+(\beta-1)\log(1-\rho)
\nonumber\\
&\quad
-\log\mathcal{B}(\alpha,\beta).
\nonumber
\end{align}
Write $\hat{p}=p(\rho;\hat{\alpha}_b,\hat{\beta}_b)$,
$p^{\star}=p(\rho;\alpha_b^{\star},\beta_b^{\star})$,
$\delta^{\alpha}_b=\hat{\alpha}_b-\alpha_b^{\star}$ and
set $\delta^{\beta}_b=\hat{\beta}_b-\beta_b^{\star}$. Subtracting the two
log-densities and simplifying by substituting $\delta^{\alpha}_b, \delta^{\beta}_b,$ we get
\begin{align}
\log\frac{\hat{p}(\rho)}{p^{\star}(\rho)}
&=
(\hat{\alpha}_b-\alpha_b^{\star})\log\rho
+(\hat{\beta}_b-\beta_b^{\star})\log(1-\rho)
\nonumber\\
&\quad
-\log\mathcal{B}(\hat{\alpha}_b,\hat{\beta}_b)
+\log\mathcal{B}(\alpha_b^{\star},\beta_b^{\star})
\nonumber\\
&=
\delta^{\alpha}_b\log\rho
+\delta^{\beta}_b\log(1-\rho)
+\log\frac{\mathcal{B}(\alpha_b^{\star},\beta_b^{\star})}
           {\mathcal{B}(\hat{\alpha}_b,\hat{\beta}_b)}.
\nonumber
\end{align}
The last term does not depend on $\rho$, so taking $\mathbb{E}_{\hat{p}}$ leaves
\begin{align}
D^{\rho}_b
&=
\delta^{\alpha}_b\,\mathbb{E}_{\hat{p}}[\log\rho]
+\delta^{\beta}_b\,\mathbb{E}_{\hat{p}}[\log(1-\rho)]
+\log\frac{\mathcal{B}(\alpha_b^{\star},\beta_b^{\star})}
           {\mathcal{B}(\hat{\alpha}_b,\hat{\beta}_b)}.
\label{eq:beta_kl_expect}
\end{align}

The only remaining integrals are the two expectations.
We simplify their expressions generally, and then apply the relevant candidate parameters. 
Starting with,
\begin{align}
\mathbb{E}[\log\rho]
&=\int_0^1\log\rho\;\frac{\rho^{\alpha-1}(1-\rho)^{\beta-1}}{\mathcal{B}(\alpha,\beta)}\,d\rho
\nonumber\\
&=\frac{1}{\mathcal{B}(\alpha,\beta)}\int_0^1
\underbrace{\rho^{\alpha-1}\log\rho}_{=\,\partial_\alpha\rho^{\alpha-1}}\,
(1-\rho)^{\beta-1}\,d\rho
\nonumber\\
&=\frac{1}{\mathcal{B}(\alpha,\beta)}\int_0^1
\frac{\partial}{\partial\alpha}\Big[\rho^{\alpha-1}(1-\rho)^{\beta-1}\Big]\,d\rho
\nonumber\\
&=\frac{1}{\mathcal{B}(\alpha,\beta)}\,\frac{\partial}{\partial\alpha}
\int_0^1\rho^{\alpha-1}(1-\rho)^{\beta-1}\,d\rho
\nonumber\\
&=\frac{1}{\mathcal{B}(\alpha,\beta)}\,\frac{\partial\mathcal{B}(\alpha,\beta)}{\partial\alpha} \nonumber \\
&=\frac{\partial}{\partial\alpha}\log\mathcal{B}(\alpha,\beta) \nonumber
\end{align}

\noindent Using $\log\mathcal{B}(\alpha,\beta)
=\log\Gamma(\alpha)+\log\Gamma(\beta)-\log\Gamma(\alpha+\beta)$ and the
digamma function $\psi(x)=\tfrac{d}{dx}\log\Gamma(x)$,
\begin{align}
\mathbb{E}[\log\rho]
&=\psi(\alpha)-\psi(\alpha+\beta)=\psi(\alpha)-\psi(\nu).
\nonumber
\end{align}
Applying the same approach but with $\partial_{\beta}(1-\rho)^{\beta-1}=(1-\rho)^{\beta-1}\log(1-\rho)$
gives the second expectation in \eqref{eq:beta_kl_expect}. That is, 
\begin{align}
\mathbb{E}[\log(1-\rho)]
&=\int_0^1 \log(1-\rho)\;\frac{\rho^{\alpha-1}(1-\rho)^{\beta-1}}{\mathcal{B}(\alpha,\beta)}\,d\rho
\nonumber\\
&=\frac{1}{\mathcal{B}(\alpha,\beta)}\int_0^1
\rho^{\alpha-1}\,
\underbrace{(1-\rho)^{\beta-1}\log(1-\rho)}_{=\,\partial_\beta(1-\rho)^{\beta-1}}\,d\rho
\nonumber\\
&=\frac{1}{\mathcal{B}(\alpha,\beta)}\int_0^1
\frac{\partial}{\partial\beta}\Big[\rho^{\alpha-1}(1-\rho)^{\beta-1}\Big]\,d\rho
\nonumber\\
&=\frac{1}{\mathcal{B}(\alpha,\beta)}\,\frac{\partial}{\partial\beta}
\int_0^1\rho^{\alpha-1}(1-\rho)^{\beta-1}\,d\rho
\nonumber\\
&=\frac{1}{\mathcal{B}(\alpha,\beta)}\,\frac{\partial\mathcal{B}(\alpha,\beta)}{\partial\beta}
\nonumber\\
&=\frac{\partial}{\partial\beta}\log\mathcal{B}(\alpha,\beta)
\nonumber\\
&=\frac{\partial}{\partial\beta}\Big[\log\Gamma(\alpha)+\log\Gamma(\beta)-\log\Gamma(\alpha+\beta)\Big]
\nonumber\\
&=\psi(\beta)-\psi(\alpha+\beta)=\psi(\beta)-\psi(\nu).
\nonumber
\end{align}

\noindent Together, for the two expectations we get:
\begin{align}
\mathbb{E}[\log\rho]&=\psi(\alpha)-\psi(\nu),
\nonumber\\
\mathbb{E}[\log(1-\rho)]&=\psi(\beta)-\psi(\nu).
\label{eq:digamma}
\end{align}
The expectations in \eqref{eq:beta_kl_expect} are under $\hat{p}$, so we
evaluate \eqref{eq:digamma} at the candidate parameters
$(\hat{\alpha}_b,\hat{\beta}_b,\hat{\nu}_b)$:
\begin{align}
D^{\rho}_b
&=
\log\frac{\mathcal{B}(\alpha_b^{\star},\beta_b^{\star})}
         {\mathcal{B}(\hat{\alpha}_b,\hat{\beta}_b)}
+\delta^{\alpha}_b\big[\psi(\hat{\alpha}_b)-\psi(\hat{\nu}_b)\big]
\nonumber\\
&\quad
+\delta^{\beta}_b\big[\psi(\hat{\beta}_b)-\psi(\hat{\nu}_b)\big].
\label{eq:beta_kl}
\end{align}
Expanding $\mathcal{B}$ with \eqref{eq:beta_fn} gives the form that we have implemented, which needs only $\log\Gamma$ and $\psi$:
\begin{align}
D^{\rho}_b
&=
\log\Gamma(\alpha_b^{\star})+\log\Gamma(\beta_b^{\star})-\log\Gamma(\nu_b^{\star})
\nonumber\\
&\quad
-\log\Gamma(\hat{\alpha}_b)-\log\Gamma(\hat{\beta}_b)+\log\Gamma(\hat{\nu}_b)
\nonumber\\
&\quad
+\delta^{\alpha}_b\,\psi(\hat{\alpha}_b)
+\delta^{\beta}_b\,\psi(\hat{\beta}_b)
-(\delta^{\alpha}_b+\delta^{\beta}_b)\,\psi(\hat{\nu}_b)
\label{eq:final_reward_implementation}
\end{align}
This completes the derivation for the expressions used in the implementation of the spatial reward in Eq. \ref{eq:spatial_reward}.
The chain of transformations starting with a latent vector $\bz$
{
\begin{align}
\bz
&\xrightarrow{\varphi}\hat\bx_0
\xrightarrow{\mathrm{STFT}}(X_L,X_R)
\xrightarrow{\mathbf{B}}(E^{LL},E^{RR},C)
\xrightarrow{\eqref{eq:iid_iic}}(\rho,\Delta)
\nonumber\\
&\xrightarrow{\text{moments}}(m,v,\hat{\mu},\hat{\sigma})
\rightarrow(\hat{\alpha},\hat{\beta})
\xrightarrow{\eqref{eq:gauss_kl},\eqref{eq:beta_kl}}r_{\mathrm{sp}}
\nonumber
\end{align}}%
uses only a neural network, linear maps (STFT, banding), products, $\log$, sample means and variances, and
$\log\Gamma$ and $\psi$. 
All are differentiable wherever their arguments are
positive, which we guarantee by applying a small $\epsilon$ in each denominator and logarithm. 
Autograd therefore differentiates \ref{eq:final_reward_implementation} exactly,
with no sampling noise.

Further, although the terms in Eq. \ref{eq:final_reward_implementation} look complex, it can be understood intuitively. 
$D^{\Delta}_b$ penalizes a wrong mean panning, measured in units
of the audio reference's own stereo spread, and a wrong amount of ILD fluctuation.
On the other hand, $D^{\rho}_b$ compares the (generated) candidate's typical coherence
$\psi(\hat\alpha_b)-\psi(\hat\nu_b)=\mathbb{E}_{\hat p}[\log\rho]$ and
decorrelation $\psi(\hat\beta_b)-\psi(\hat\nu_b)=\mathbb{E}_{\hat p}[\log(1-\rho)]$
with the target's, weighted by the parameter errors $\delta^\alpha_b,\delta^\beta_b$.
Both vanish only when the candidate matches the reference, and errors cost
more in bands where the reference is stable, so $r_{\mathrm{sp}}$ rewards
matching the reference's width and its temporal stability, not only its mean.

\subsection{Inner Optimization Schedule}
\label{app:perf_impr}
\noindent
The Stable Audio 3 model is post-trained for a fixed schedule
$1>t_0>t_1>\dots>t_{N-1}>t_N=0$ with $N=8$. Querying it at other noise levels showed degradations in its estimates leading to audible artifacts in the generated music. 
Therefore, instead of adding new intermediate noise levels, we revisit the ones used in post-training as a way to increase optimization iterations. 
After the two phases at step $i$, the result is re-noised either
to the same level, $x\leftarrow(1-t_i)z^{\star}+t_i\varepsilon$, or to the next
level $t_{i+1}$. 
Every model call therefore still happens at one of the $N$ trained noise levels.

A re-noise to $t_i$ keeps a fraction $1-t_i$ of the optimized latent. 
At high noise, a repeat mostly discards the refinement and behaves like a restart. 
Near
the data manifold ($t_i\to0$), it keeps most of the refinement, so successive
passes build on each other. 
We therefore split a budget of $R$ passes in proportion to this retention:
{
\footnotesize
\begin{align}
r_i=\max\!\left(1,\ \operatorname{round}\!\left(
R\,\frac{1-t_i}{\sum_{j=0}^{N-1}(1-t_j)}\right)\right),
\quad i=0,\dots,N-1 \nonumber
\label{eq:repeats}
\end{align}}%
Step $i$ is visited $r_i$ times, so the sampler runs on the expanded schedule given as,
{
\begin{align}
\big(
\underbrace{t_0,\dots,t_0}_{r_0},\
\underbrace{t_1,\dots,t_1}_{r_1},\
\dots,\
\underbrace{t_{N-1},\dots,t_{N-1}}_{r_{N-1}},\
t_N
\big)
\nonumber
\end{align}}%
Each visit costs one model evaluation plus $K_1+K_2$ decoder
forward--backward passes. 
The total is $\sum_i r_i\approx R$ model evaluations (rounding can shift it slightly), and most of them are spent at low noise,
where the spatial cues of the decoded audio are already well defined and the spatial-reward gradient can act on them.

\end{document}